\documentclass[11pt,a4paper]{article}
\pdfoutput=1

\usepackage{jcappub}

\usepackage{amssymb,amsmath,tensor,slashed}
\usepackage{graphicx}

\newcommand{\dd}[1]{\mathrm{d}#1\,}

\begin{document}
\title{Gravitationally bound BCS state as dark matter}
\author[a]{Stephon Alexander}
\author[b]{Sam Cormack}
\affiliation[a]{Department of Physics, Brown University\\ Providence, RI 20912}
\affiliation[b]{Department of Physics and Astronomy, Dartmouth College\\ Hanover, NH 03755}

\emailAdd{stephon\_alexander@brown.edu}
\emailAdd{samuel.c.cormack.gr@dartmouth.edu}

\abstract{
We explore the possibility that fermionic dark matter undergoes a BCS transition to form a superfluid. This requires an attractive interaction between fermions and we describe a possible source of this interaction induced by torsion. We describe the gravitating fermion system with the Bogoliubov-de Gennes formalism in the local density approximation. We solve the Poisson equation along with the equations for the density and gap energy of the fermions to find a self-gravitating, superfluid solution for dark matter halos. In order to produce halos the size of dwarf galaxies, we require a particle mass of $\sim 200\mathrm{eV}$. We find a maximum attractive coupling strength before the halo becomes unstable. If dark matter halos do have a superfluid component, this raises the possibility that they contain vortex lines.}

\arxivnumber{1607.08621}

\maketitle

\section{Introduction}
There has recently been a significant amount of interest in the possibility that dark matter forms a many-body quantum state in galactic halos. Much of this work has focused on bosonic dark matter \cite{Sin:1994,Ji:1994,Peebles:2000,Goodman:2000,Hu:2000,Boehmer:2007,Harko:2011,Chavanis:2011,Harko:2011a,Harko:2011b,
RindlerDaller:2011,Pires:2012,Bettoni:2013,Li:2013,Diez-Tejedor:2014,Fan:2016}. A massive bosonic particle can undergo Bose-Einstein condensation and be supported by its own gravitational interactions. One of the reasons for the popularity of this model is that it provides a possible solution to the core-cusp problem of dark matter \cite{deBlok:2009}. This problem arises from large-scale simulations of dark matter structure formation which predict large cuspy densities in the centers of galaxies. These density profiles do not match observations of the galaxies. In Bose-Einstein condensates these cusps cannot form due to either repulsive interactions, or the uncertainty principle in the case of non- or weakly interacting bosons. Recent work has explored the idea that Bose condensation of dark matter and coupling to baryons could explain the MOND-like behaviour of rotation curves while keeping the large-scale success of particle dark matter \cite{Berezhiani:2015,Berezhiani:2016}. The formation of a Bose-Einstein condensate leads to the dark matter halo behaving as a superfluid.

The possibility that fermionic dark matter could form a many-body quantum state has also been explored \cite{Chavanis:2002,deVega:2013,Destri:2013,Chavanis:2015,Domcke:2015}. Previous work has focused on the idea that fermion dark matter can form a degenerate Fermi gas either in the core or throughout the whole of a dwarf galaxy halo. These models also avoid the core-cusp problem as the degeneracy pressure stops large overdensities from building up. Unlike the bosonic case, a degenerate Fermi gas does not behave as a superfluid, however as we will see, adding an attractive interaction allows a superfluid to form via a Bardeen-Cooper-Schrieffer (BCS) transition.

One significant difference between the degenerate fermion and boson models is the particle masses required to produce galactic sized haloes supported by quantum or degeneracy pressure. On dimensional grounds, the size of a bosonic halo supported by quantum pressure can be estimated to be \cite{Domcke:2015}
\begin{equation}
R\sim \frac{h^2}{GMm_b^2}
\end{equation}
where $m_b$ is the mass of the boson and $M$ the mass of the halo. For a fermionic halo supported by degeneracy pressure, the size will be
\begin{equation}
\label{eq:fermiMR}
R\sim \left(\frac{M}{m_f}\right)^{2/3}\frac{h^2}{GMm_f^2}
\end{equation}
with $m_f$ the mass of the fermion. Since $M\gg m_f$ the halo radius will be much larger for a fermion of a given mass than a boson with the same mass. As an estimate for Milky Way sized galaxy, Domcke and Urbano \cite{Domcke:2015} take $R\sim 100\mathrm{kpc}$ and $M=10^{12}M_\odot$ giving a boson with $m_b\sim 10^{-25}\mathrm{eV}$ and a fermion with $m_f\sim 20\mathrm{eV}$. It should be noted that larger boson masses can be allowed if the condensate is supported primarily by a repulsive interaction rather than the quantum pressure, and in fact non-interacting bosonic dark matter is disfavored by CMB measurements \cite{Li:2013}.

Here we investigate the possibility that a fermionic species can become a superfluid via a BCS transition. Fermions can undergo a transition to a superfluid state via the BCS mechanism if there is at least a small attractive force between particles. We model a self-gravitating system of fermions with an attractive self-coupling. We find self-consistent solutions for the fermion density and gap energy along with the gravitational potential.

\section{Model}
We consider a system of a single fermionic species with an attractive self-interaction coupled to gravity, with the action
\begin{equation}
S = \int\dd{^4x}\sqrt{-g}\left[\mathcal{L}_\mathrm{Grav} +\mathcal{L}_\psi \right]
\end{equation}
with the usual Einstein-Hilbert gravitational Lagrangian,
\begin{equation}
\mathcal{L}_\mathrm{Grav}=\frac{c^4}{16\pi G} R,
\end{equation}
and the fermion Lagrangian,
\begin{equation}
\mathcal{L}_\psi = -\bar{\psi}_D\left(i\hbar c\gamma^\mu\partial_\mu+mc^2\right)\psi_D+\frac{g}{2}(\bar{\psi}_D\psi_D)^2,
\end{equation}
where $g$ is a (dimensionful) self-coupling constant for the fermion. If $g>0$, the interaction is attractive. We will take an effective field theory view of the interaction where we remain agnostic about the source of the coupling. As long as the energy scale of the dynamics remain below the mass scale implied by the coupling ($M^*\sim g^{-1/2}$), this will be valid. Such an interaction may be generated by a Yukawa coupling to a heavy `dark' scalar field where the mass of the scalar field is $M^*$. Alternatively, a four-fermion interaction may be generated by torsion. The torsion-mediated interaction may be made attractive by including a Holst term in the gravitational action with an imaginary Immirzi parameter \cite{Freidel:2005,Perez:2006}. The case of an imaginary Immirzi parameter leads to the Ashtekar self-dual formalism for quantum gravity \cite{Ashtekar:2004}.

We will work in the non-relativistic, weak field regime so we assume the metric takes the form, $g_{\mu\nu} = \mathrm{diag}(-(1+2\phi/c^2),\delta_{ij}(1-2\phi/c^2))$. The equation of motion for the gravitational field is then just the Poisson equation,
\begin{equation}
\nabla^2\phi = 4\pi G T_{00}
\end{equation}
where $T_{00}$ is the energy density of the fermions. In the weak field, non relativistic case, the dominant term is
\begin{equation}
T_{00} = mc^2\bar{\psi}_D\gamma^0\psi_D = mc^2\psi_D^\dagger\psi_D.
\end{equation}
In order to take the non relativistic limit for the fermions, we use the Dirac representation of the gamma matrices and write $\psi_D$ as
\begin{equation}
\psi_D = \left(\begin{array}{c}
\psi \\
\chi\\
\end{array}\right)
\end{equation}
where $\psi$ and $\chi$ are two-component spinors. The Dirac equation reduces to  a Schr\"odinger equation for $\psi$ with a Hamiltonian,
\begin{equation}
H_\mathrm{sp} = -\frac{\hbar^2}{2m}\nabla^2+m\phi
\label{eq:H0}
\end{equation}
while $\chi$ is related to $\psi$ by
\begin{equation}
\chi = \frac{-i\hbar}{2mc}\sigma^i\partial_i\psi \ll \psi
\end{equation}
Applying the same decomposition to the interaction term, and dropping terms involving $\chi$ we get an interaction potential
\begin{equation}
V_\mathrm{int} = -\frac{g}{2}\psi^\dagger\psi\psi^\dagger\psi
\end{equation}
If we write $\psi$ in terms of explicit components, $\psi=(\psi_\downarrow,\psi_\uparrow)$, and treat the components as fermionic operators, the interaction can be written as
\begin{equation}
V_\mathrm{int} = -g\psi_\downarrow^\dagger\psi_\uparrow^\dagger\psi_\uparrow\psi_\downarrow.
\label{eq:Vint}
\end{equation}

\section{Formalism}
In order to study an inhomogeneous BCS state we use the Bogoliubov-de Gennes formalism \cite{DeGennes:1966,Tinkham:1996}. We use the single particle Hamiltonian, \eqref{eq:H0}, and the contact interaction, \eqref{eq:Vint}, to write down a non-relativistic many particle Hamiltonian,
\begin{equation}
\label{eq:hamiltonian}
H = \int\dd{^3x} \sum_{i=\uparrow,\downarrow}{\left[\psi_i^\dagger(\mathbf{x}) H_\mathrm{sp} \psi_i(\mathbf{x})\right]} - g\psi_\downarrow^\dagger(\mathbf{x})\psi_\uparrow^\dagger(\mathbf{x})\psi_\uparrow(\mathbf{x})\psi_\downarrow(\mathbf{x})
\end{equation}
where $\psi_i$ are the components of the fermionic field operators. The gravitational potential is determined from the Poisson equation,
\begin{equation}
\nabla^2\phi = 4\pi Gmc^2n(\mathbf{x})
\label{eq:poisson}
\end{equation}
where $n(\mathbf{x})$ is the total number density of fermions, $n(\mathbf{x})=\langle\psi_D^\dagger\psi_D\rangle = \langle\psi_\uparrow^\dagger(\mathbf{x})\psi_\uparrow(\mathbf{x})\rangle + \langle\psi_\downarrow^\dagger(\mathbf{x})\psi_\downarrow(\mathbf{x})\rangle$, dropping the $\chi$ terms. At this stage we are already treating the gravitational potential classically which is acceptable as long as the total particle number is large. The Poisson equation will need to be solved self-consistently along with the equations for the fermions.

We now follow standard BCS theory in treating the attractive contact interaction within mean field theory as well. We therefore make the approximation,
\begin{equation}
-g\psi_\downarrow^\dagger\psi_\uparrow^\dagger\psi_\uparrow\psi_\downarrow \approx \Delta\psi_\downarrow^\dagger\psi_\uparrow^\dagger + \Delta^*\psi_\uparrow\psi_\downarrow - \frac{g}{2}n(\mathbf{x})\psi^\dagger_\uparrow\psi_\uparrow - \frac{g}{2}n(\mathbf{x})\psi^\dagger_\downarrow\psi_\downarrow
\end{equation}
where $\Delta(\mathbf{x})$, the gap field, is defined as
\begin{equation}
\Delta(\mathbf{x}) = -g\langle\psi_\uparrow(\mathbf{x})\psi_\downarrow(\mathbf{x})\rangle
\label{eq:gappsi}
\end{equation}
and we have assumed that $\langle\psi_\uparrow^\dagger(\mathbf{x})\psi_\uparrow(\mathbf{x})\rangle =\langle\psi_\downarrow^\dagger(\mathbf{x})\psi_\downarrow(\mathbf{x})\rangle =n(\mathbf{x})/2$ in writing the Hartree mean field terms. These terms represent the bulk effect of the attractive interaction, while the terms involving the gap represent the formation of Cooper pairs.
We will work in the grand canonical ensemble so we introduce a chemical potential, $\mu$ to control the average particle number. We then have a grand canonical Hamiltonian,
\begin{equation}
K = \int\dd{^3x} \sum_{i=\uparrow,\downarrow}{\left[\psi_i^\dagger(\mathbf{x}) (H_0-\mu) \psi_i(\mathbf{x})\right]} + \Delta\psi_\downarrow^\dagger\psi_\uparrow^\dagger + \Delta^*\psi_\uparrow\psi_\downarrow
\end{equation}
where we have absorbed the Hartree terms into the single particle Hamiltonian,
\begin{equation}
H_0 = H_\mathrm{sp} -\frac{g}{2}n(\mathbf{x}).
\end{equation} 
Our aim is to find the ground state of this Hamiltonian. We will do this by applying a Bogoliubov transformation to the field operators in order to diagonalize the Hamiltonian. The ground state of the Hamiltonian is then just the vacuum state of the new field operators.

To diagonalize the Hamiltonian we first write it in terms of a matrix,
\begin{equation}
M = \left(\begin{array}{cc}
 H_0-\mu & \Delta \\
 \Delta^* & -(H_0 - \mu)\\
\end{array}\right)
\end{equation}
and vector
\begin{equation}
\Psi(\mathbf{x})=\left(\begin{array}{c}\psi_\downarrow(\mathbf{x})\\ \psi_\uparrow^\dagger(\mathbf{x}) \end{array} \right)
\end{equation}
as 
\begin{equation}
K = \int\dd{^3x}\Psi^\dagger(\mathbf{x})M\Psi(\mathbf{x}).
\label{eq:matrixK}
\end{equation}
To diagonalize $M$ we must solve the eigenvalue equation,
\begin{equation}
\left(\begin{array}{cc}
 H_0-\mu & \Delta \\
 \Delta^* & -(H_0 - \mu)\\
\end{array}\right)
\left(\begin{array}{c}
u_a(\mathbf{x})\\
v_a(\mathbf{x})\\
\end{array}\right)
= E_a \left(\begin{array}{c}
u_a(\mathbf{x})\\
v_a(\mathbf{x})\\
\end{array}\right)
\label{eq:BdG}
\end{equation}
These are the Bogoliubov-de Gennes (BdG) equations which are to be solved for the modes $u_a(\mathbf{x})$ and $v_a(\mathbf{x})$, and the eigenvalues $E_a$ \cite{DeGennes:1966}. If the equations are satisfied by $(u_a,v_a)$  with eigenvalue $E_a$ then they are also satisfied by $(-v^*_a,u^*_a)$ with eigenvalue $-E_a$. Since the two sets of solutions are not independent, we use only the positive $E_a$ solutions when calculating other quantities. The similarity transformation for the Hamiltonian is then given by the Bogoliubov transformation of the field operators,
\begin{equation}
\Psi(\mathbf{x})\equiv\left(\begin{array}{c}\psi_\downarrow(\mathbf{x})\\ \psi_\uparrow^\dagger(\mathbf{x}) \end{array} \right) = 
\sum_a{\left(\begin{array}{cc}
 u_a(\mathbf{x}) & -v^*_a(\mathbf{x}) \\
 v_a(\mathbf{x}) & u^*_a(\mathbf{x}) \\
 \end{array}\right)}
\left(\begin{array}{c}
 \alpha_a\\
 \beta_a^\dagger \\
 \end{array}\right)
\label{eq:bogTrans}
\end{equation}
where $a$ is a quantum number labeling the modes $u_a$ and $v_a$ created by the operators $\alpha_a^\dagger$ and $\beta_a^\dagger$. The transformation for $\psi_\uparrow$ and $\psi_\downarrow^\dagger$ are given by the conjugate of equation \eqref{eq:bogTrans}. The solutions of equation \eqref{eq:BdG} are orthogonal and we can choose them normalized such that the modes obey the orthogonality relation
\begin{equation}
\int\dd{^3x}
\left(\begin{array}{cc}
 u_b^*(\mathbf{x}) & v^*_b(\mathbf{x}) \\
 -v_b(\mathbf{x}) & u_b(\mathbf{x}) \\
 \end{array}\right)
\left(\begin{array}{cc}
 u_a(\mathbf{x}) & -v^*_a(\mathbf{x}) \\
 v_a(\mathbf{x}) & u^*_a(\mathbf{x}) \\
 \end{array}\right)
= \delta_{ab}\left(\begin{array}{cc}
1 & 0\\
0 & 1\\
\end{array}\right)
\end{equation}
and the $\alpha_a$ and $\beta_a$ operators obey anticommutation relations, $\lbrace \alpha_a,\alpha_b^\dagger \rbrace = \delta_{ab}$ and $\lbrace \beta_a,\beta_b^\dagger \rbrace = \delta_{ab}$. We then substitute equation \eqref{eq:bogTrans} into the Hamiltonian \eqref{eq:matrixK} and use the orthogonality relation to get, ignoring the ground state energy,
\begin{equation}
K = \sum_a{E_a\left(\alpha_a^\dagger\alpha_a + \beta_a^\dagger\beta_a\right)}
\end{equation}
so that the Hamiltonian is diagonalized in terms of the operators $\alpha_a$ and $\beta_a$. Finally, the BCS ground state $|0_\mathrm{BCS}\rangle$ is identified with the ground state of this Hamiltonian, or the vacuum state of the operators $\alpha_a$ and $\beta_a$. We will assume zero temperature in this work, so our system will always be in the BCS ground state.

The BdG equations \eqref{eq:BdG} must be solved self-consistently with an equation for the gap. To do this we rewrite the gap equation \eqref{eq:gappsi} in terms of the Bogoliubov modes. If the expectation value is taken in the state $|0_\mathrm{BCS}\rangle$, then we get
\begin{equation}
\Delta(\mathbf{x}) = g\sum_a{u_a(\mathbf{x})v^*_a(\mathbf{x})}.
\label{eq:gapbog}
\end{equation}
In order to calculate the self-consistent gravitational field we need the number density of fermions. Again, we can write this in terms of the Bogoliubov modes. At zero temperature it is
\begin{equation}
n(\mathbf{x}) = 2\sum_a |v_a(\mathbf{x})|^2.
\label{eq:densbog}
\end{equation}
In principle, we could now solve equations \eqref{eq:poisson}, \eqref{eq:BdG}, \eqref{eq:gapbog} and \eqref{eq:densbog} self-consistently, however, there are two problems with this. The first is that the equation for the gap \eqref{eq:gapbog} will turn out to be divergent due to the assumption of a contact interaction. The second is that when we have a very large number of particles, the number of eigenstates that need to be calculated is also very large.

In order to solve the issue of having to calculate too many states, we will make a semiclassical or local density approximation (LDA) \cite{Grasso:2003}. In this approximation, the system is treated as locally homogeneous This will also show how the gap equation diverges and give us a way to remove the divergence. First, we rewrite the Bogoliubov amplitudes in the form
\begin{equation}
u_a(\mathbf{x}) = u(\mathbf{x},\mathbf{k})e^{i\mathbf{k}\cdot\mathbf{x}},\:\:
v_a(\mathbf{x}) = v(\mathbf{x},\mathbf{k})e^{i\mathbf{k}\cdot\mathbf{x}}
\end{equation}
where we have replaced the quantum number $a$ with a continuous variable, $\mathbf{k}$ so that we are treating the modes as plane waves with varying amplitude. We are essentially making a continuum approximation for the energy levels which will be valid when there are many particles present to occupy states. The single particle Hamiltonian acts on these states to give
\begin{multline}
H_0\left[u(\mathbf{x},\mathbf{k})e^{i\mathbf{k}\cdot\mathbf{x}}\right] = (m\phi(\mathbf{x})-\frac{g}{2}n(\mathbf{x}))u(\mathbf{x},\mathbf{k})e^{i\mathbf{k}\cdot\mathbf{x}} \\
+\frac{\hbar^2}{2m}\left[k^2u(\mathbf{x},\mathbf{k})-2i\mathbf{k}\cdot\nabla u(\mathbf{x},\mathbf{k}) -\nabla^2 u(\mathbf{x},\mathbf{k})\right]e^{i\mathbf{k}\cdot\mathbf{x}}
\end{multline}
The local density approximation amounts to dropping the terms involving derivatives of $u(\mathbf{x},\mathbf{k})$. We can think of this as assuming the wavelength of the particles is much shorter than the scale on which their amplitude varies. There will always be some states with $\mathbf{k}<\nabla u$ since, in this quasicontinuum limit there is no limit on how small $\mathbf{k}$ can be. However, if $\mathbf{k}_F\gg \nabla u$, where $\mathbf{k}_F$ is the Fermi wavevector, and there are many states below the Fermi level, then the error in the low lying states will be irrelevant. In terms of the fermion density and chemical potential, this requirement can be expressed as
\begin{equation}
\frac{\nabla n}{n} \ll k_F(\mathbf{x})
\end{equation}
where
\begin{equation}
k_F(\mathbf{x}) \equiv \frac{\sqrt{2m}}{\hbar}\sqrt{\mu-m\phi(\mathbf{x})+\frac{g}{2}n(\mathbf{x})}
\label{eq:kFermi}
\end{equation}
is the local Fermi wavevector. This inequality will tend to break down around the edges of the BCS condensate.

In the local density approximation the BdG equation \eqref{eq:BdG} becomes an algebraic eigenvalue equation
\begin{equation}
\left(\begin{array}{cc}
\xi(\mathbf{x},\mathbf{k})-\mu & \Delta(\mathbf{x}) \\
\Delta^*(\mathbf{x}) & -(\xi(\mathbf{x},\mathbf{k})-\mu)\\
\end{array}\right)
\left(\begin{array}{c}
u(\mathbf{x},\mathbf{k})\\
v(\mathbf{x},\mathbf{k})\\
\end{array}\right)
= E(\mathbf{x},\mathbf{k})
\left(\begin{array}{c}
u(\mathbf{x},\mathbf{k})\\
v(\mathbf{x},\mathbf{k})\\
\end{array}\right)
\end{equation}
where 
\begin{equation}
\xi(\mathbf{x},\mathbf{k}) = \frac{\hbar^2k^2}{2m}+m\phi(\mathbf{x})-\frac{g}{2}n(\mathbf{x}).
\end{equation}
The solutions are simply
\begin{equation}
\begin{gathered}
|u(\mathbf{x},\mathbf{k})|^2 = \frac{1}{2}\left(1+\frac{\xi(\mathbf{x},\mathbf{k})-\mu}{E(\mathbf{x},\mathbf{k})}\right)\\
|v(\mathbf{x},\mathbf{k})|^2 = \frac{1}{2}\left(1-\frac{\xi(\mathbf{x},\mathbf{k})-\mu}{E(\mathbf{x},\mathbf{k})}\right)\\
u(\mathbf{x},\mathbf{k})v^*(\mathbf{x},\mathbf{k}) = \frac{\Delta(\mathbf{x})}{2E(\mathbf{x},\mathbf{k})}\\
E(\mathbf{x},\mathbf{k}) = \frac{\hbar^2}{2m}\sqrt{(k^2-k_F(\mathbf{x})^2)^2+4\frac{m^2}{\hbar^4}|\Delta(\mathbf{x})|^2}
\end{gathered}
\end{equation}
with the self-consistency condition for the gap given by
\begin{equation}
\frac{1}{g} = \int\frac{\dd{^3k}}{(2\pi)^3}\frac{1}{2E(\mathbf{x},\mathbf{k})}.
\label{eq:gapLDA}
\end{equation}
The fermion density is now given by
\begin{equation}
\label{eq:densLDA}
n(\mathbf{x}) = \int\frac{\dd{^3k}}{(2\pi)^3}\left(1 - \frac{\xi(\mathbf{x},\mathbf{k})-\mu}{E(\mathbf{x},\mathbf{k})} \right).
\end{equation}

The integral \eqref{eq:gapLDA} for the gap does not converge for large $k$, just as for a homogeneous superconductor with a contact interaction. In that case, a cutoff at the Debye energy is used since the interaction is mediated by phonons. In the present case there is no obvious cutoff to use. Instead we can apply a regularization approach from cold atom physics \cite{Bulgac:2002,Grasso:2003}. We can then write a regularized equation for the gap,
\begin{equation}
\label{eq:reggap}
\frac{1}{g} = \int_{|k|<k_c}\frac{\dd{^3k}}{(2\pi)^3}\frac{1}{2E(\mathbf{x},\mathbf{k})} 
-\frac{mk_c(\mathbf{x})}{2\pi^2\hbar^2}\left[1-\frac{k_F(\mathbf{x})}{2k_c(\mathbf{x})}\ln\frac{k_c(\mathbf{x})+k_F(\mathbf{x})}{k_c(\mathbf{x})-k_F(\mathbf{x})}\right]
\end{equation}
where $k_c(\mathbf{x})$ is a momentum cutoff defined for a constant energy cutoff, $E_c$ as
\begin{equation}
\frac{\hbar^2k_c^2(\mathbf{x})}{2m} = E_c+\mu+\frac{g}{2}n(\mathbf{x})-m\phi(\mathbf{x}).
\end{equation}
The right hand side of equation \eqref{eq:reggap} converges as $k_c\rightarrow\infty$, however for any numerical work we must use some finite value of $E_c\gg\mu$.

Equation \eqref{eq:reggap} still presents numerical issues as the integral diverges for $\Delta(\mathbf{x})=0$ due to the singularity at $k=k_F$. There will always be regions where the gap is small and the integral will be difficult to integrate numerically there. It is therefore important to have an analytical approximation for the integral
\begin{equation}
I = \int_{|k|<k_c}\frac{\dd{^3k}}{(2\pi)^3}\frac{1}{2E(\mathbf{x},\mathbf{k})}
= \frac{m}{2\pi^2\hbar^2}\int_0^{k_c}\dd{k}\frac{k^2}{\sqrt{(k^2-k_F(\mathbf{x})^2)^2+4\frac{m^2}{\hbar^4}|\Delta(\mathbf{x})|^2}}
\end{equation}
 when the gap is small. We expect that the integral will be dominated by the contribution near $k=k_F$ so we split the integral into three parts
 \begin{equation}
 I = I_1 +I_2+I_3 = \int_0^{k_F-\epsilon}+\int_{k_F-\epsilon}^{k_F+\epsilon}+\int_{k_F+\epsilon}^{k_c}
 \end{equation}
 where $I_1$, $I_2$ and $I_3$ refer to the three integrals respectively. For $\epsilon\ll k_F$, the second integral can be evaluated to leading order giving
 \begin{equation}
 I_2 \approx \frac{m}{2\pi^2\hbar^2}\frac{k_F}{2}\ln\frac{k_F\epsilon+\sqrt{k_F^2\epsilon^2+m^2|\Delta|^2/\hbar^4}}{-k_F\epsilon+\sqrt{k_F^2\epsilon^2+m^2|\Delta|^2/\hbar^4}}
 \end{equation}
For the rest of the integral, we can safely Taylor expand in $m^2|\Delta|^2/(\hbar^4(k^2-k_F^2)^2)$ as long as $m|\Delta|/(\hbar^2k_F\epsilon) \ll 1$. The lowest order contribution in this expansion is
\begin{equation}
I_1+I_2 = \frac{mk_c(\mathbf{x})}{2\pi^2\hbar^2}\left[1-\frac{k_F(\mathbf{x})}{2k_c(\mathbf{x})}\ln\frac{k_c(\mathbf{x})+k_F(\mathbf{x})}{k_c(\mathbf{x})-k_F(\mathbf{x})}\right]
\end{equation}
which is precisely the divergent term cancelled by the regularization. Combining the two requirements on $\epsilon$ gives
\begin{equation}
\frac{m|\Delta|}{\hbar^2k_F}\ll\epsilon\ll k_F
\end{equation}
so the approximation will only be valid for $|\Delta|\ll\hbar^2k_F^2/m$. Using the requirement on $|\Delta|$ in the expression for $I_2$ we get
\begin{equation}
I_2 = \frac{-mk_F}{2\pi^2\hbar^2}\ln\frac{m|\Delta|}{\hbar^2k_F^2}
\end{equation}
The non-divergent parts of $I_1$ and $I_3$ are higher order in $m|\Delta|/\hbar^2k_F\epsilon$ so this expression for $I_2$ is a good approximation for $I$, useful for numerics when the gap is small. The gap equation can then be solved analytically for small $|\Delta|$, giving
\begin{equation}
\label{eq:gapsmall}
|\Delta(\mathbf{x})| = \frac{\hbar^2k_F^2(\mathbf{x})}{m}\exp\left(-\frac{2\pi^2\hbar^2}{mgk_F(\mathbf{x})} \right).
\end{equation}

We can also calculate the leading order corrections to the energy density which sources the Poisson equation \eqref{eq:poisson}. To do this we take the expectation value of the integrand of equation \eqref{eq:hamiltonian} which is the Hamiltonian density. This gives a correction to the energy density,
\begin{equation}
\delta\rho = E_K(\mathbf{x})+mn(\mathbf{x})\phi(\mathbf{x})-gn(\mathbf{x})^2-2\frac{|\Delta(\mathbf{x})|^2}{g}
\end{equation}
where
\begin{equation}
\label{eq:rhocorr}
E_K(\mathbf{x}) = \frac{\hbar^2}{2m}\int\frac{\dd{^3k}}{(2\pi)^3}\frac{k^2}{2}\left(1 - \frac{\xi(\mathbf{x},\mathbf{k})-\mu}{E(\mathbf{x},\mathbf{k})} \right)
\end{equation}
is the kinetic energy density.

\section{Solutions}
For these solutions we will assume the dark matter halo to be spherically symmetric. First consider the case where the fermions are non-interacting, $g=0$. There are two  parameters to vary: the particle mass $m$ and the chemical potential $\mu$. These parameters can be absorbed into dimensionless variables leaving a system of equations with no adjustable parameters. In this section we also use units where $\hbar=c=1$. As suitable choice of dimensionless variables is 
\begin{equation}
\label{eq:dimless}
\begin{gathered}
\tilde{n} \equiv \nu^{3/2}\frac{n}{|\mu|^{3/2}m^{3/2}}\\
\tilde{k} \equiv \nu^{1/2}\frac{k}{|\mu|^{1/2}m^{1/2}}\\
\tilde{r} \equiv \nu^{-1/4}\frac{|\mu|^{1/4}m^{7/4}}{M_p}r\\
\tilde{\phi} \equiv \nu\frac{m}{|\mu|}\phi
\end{gathered}
\end{equation}
where tildes indicate the dimensionless variables ($\phi$ is already dimensionless, so $\tilde{\phi}$ is just rescaled) and $\nu$ is an arbitrary dimensionless positive scaling factor. In principle $\nu$ can be set to one, however it turns out to be useful to allow it to vary in finding numerical solutions. With the usual boundary condition, $\phi\rightarrow 0$ as $r\rightarrow\infty$, we need the chemical potential, $\mu$ to be negative so that the density vanishes at some radius. It should also be greater than the minimum of $m\phi(r)$ so that the density does not vanish everywhere.

In the non-interacting case, equation \eqref{eq:densLDA} for the density can be solved analytically in terms of the gravitational potential giving
\begin{equation}
\label{eq:densg0}
\tilde{n}(\tilde{r}) = \frac{1}{3\pi^2}\left(-2\nu-2\tilde{\phi}(\tilde{r})\right)^{3/2}
\end{equation}
while the Poisson equation \eqref{eq:poisson} becomes
\begin{equation}
\label{eq:phig0}
\frac{\mathrm{d}^2}{\mathrm{d}\tilde{r}^2}\left(\tilde{r}\tilde{\phi}(\tilde{r})\right) = \frac{1}{2}\tilde{r}\tilde{n}(\tilde{r})
\end{equation}

We can re-derive the mass-radius relation of equation \eqref{eq:fermiMR} using the rescaled variables. The halo mass is given by
\begin{equation}
M = \int\dd{^3x}mn(\mathbf{x}).
\end{equation}
Using the dimensionless variables in this equation and defining $r_0\equiv M_p/\mu^{1/4}m^{7/4}$ as the natural radial unit implied by the non-dimensionalization, we get
\begin{equation}
r_0 = \frac{M_p^2\tilde{N}^{1/3}}{m^{8/3}M^{1/3}}
\end{equation}
where $\tilde{N}\equiv \int\dd{^3\tilde{x}}\tilde{n}$. This agrees with the mass-radius relation determined by equating the degeneracy pressure and gravitational attraction \cite{Domcke:2015}.

\begin{figure}
\includegraphics[scale=1]{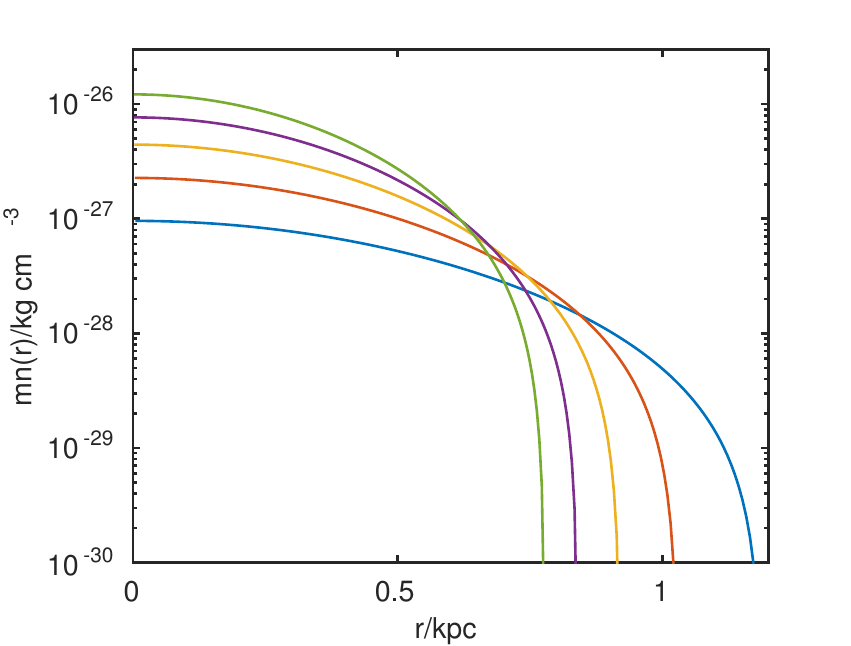}
\caption{\label{fig:densg0}Density profile for non-interacting degenerate fermions. The fermion mass is taken to be $m=200\mathrm{eV}$. The different profiles correspond to different chemical potentials, $|\mu|=0.13\mathrm{\mu eV}$ (blue), $0.24\mathrm{\mu eV}$ (red), $0.37\mathrm{\mu eV}$ (yellow), $0.53\mathrm{\mu eV}$ (purple), $0.73\mathrm{\mu eV}$ (green).}
\end{figure}

We solve equations \eqref{eq:densg0} and \eqref{eq:phig0} iteratively until a self consistent solution for $\tilde{n}$ and $\tilde{\phi}$ are found. At each step, $\nu$ is adjusted such that $\nu-\tilde{\phi}(0)=1$. If $\nu$ is held fixed, the iteration does not converge to the solution. By holding $\nu-\tilde{\phi}(0)$ fixed instead, the iteration converges for some eventual value of $\nu$. This value of $\nu$ is then used to rescale the variables. In figure \ref{fig:densg0} we reproduce figure 2 of the work by Domcke and Urbano \cite{Domcke:2015} using our formalism.

We now introduce the attractive coupling between fermions which adds the coupling  parameter, $g$, to the problem. We define dimensionless variables for the coupling constant and gap energy,
\begin{equation}
\begin{gathered}
\tilde{g} \equiv \nu^{-1/2}|\mu|^{1/2}m^{3/2}g\\
\tilde{\Delta} = \frac{\nu}{|\mu|}\Delta.
\end{gathered}
\end{equation}
The self-consistent set of dimensionless equations to be solved for the fermion density $\tilde{n}$, gravitational potential $\tilde{\phi}$ and gap energy $\tilde{\Delta}$ are respectively
\begin{gather}
\tilde{n}(\tilde{r}) = \frac{1}{2\pi^2}\int_0^{\tilde{k}_c}\dd{\tilde{k}}\tilde{k}^2\left[1-\frac{\tilde{k}^2-\tilde{k}_F(\tilde{r})^2}{\sqrt{(\tilde{k}^2-\tilde{k}_F(\tilde{r})^2)^2+4\tilde{\Delta}(\tilde{r})^2)}}\right]\\
\frac{\mathrm{d}^2}{\mathrm{d}\tilde{r}^2}\left(\tilde{r}\tilde{\phi}(\tilde{r})\right) = \frac{1}{2}\tilde{r}\tilde{n}(\tilde{r})\\
\frac{2\pi^2}{\tilde{g}}+\tilde{k}_c\left[1-\frac{\tilde{k}_F}{2\tilde{k}_c}\log\left(\frac{\tilde{k}_c+\tilde{k}_F}{\tilde{k}_c-\tilde{k}_F}\right)\right] = \int_0^{\tilde{k}_c} \dd{\tilde{k}}\frac{\tilde{k}^2}{\sqrt{(\tilde{k}^2-\tilde{k}_F^2)^2+4\tilde{\Delta}^2}}
\end{gather}
where the dimensionless Fermi and cutoff wavevectors are
\begin{gather}
\tilde{k}_F^2 = 2\left[-\nu+\frac{1}{2}\tilde{g}\tilde{n}-\tilde{\phi}\right]\\
\tilde{k}_c^2 = 2\left[\nu\left(\frac{E_c}{|\mu|}-1\right)+\frac{1}{2}\tilde{g}\tilde{n}-\tilde{\phi} \right]
\end{gather}
and we should choose $E_c\gg|\mu|$ in order for the integrals to converge.

\begin{figure}
\includegraphics[scale=1]{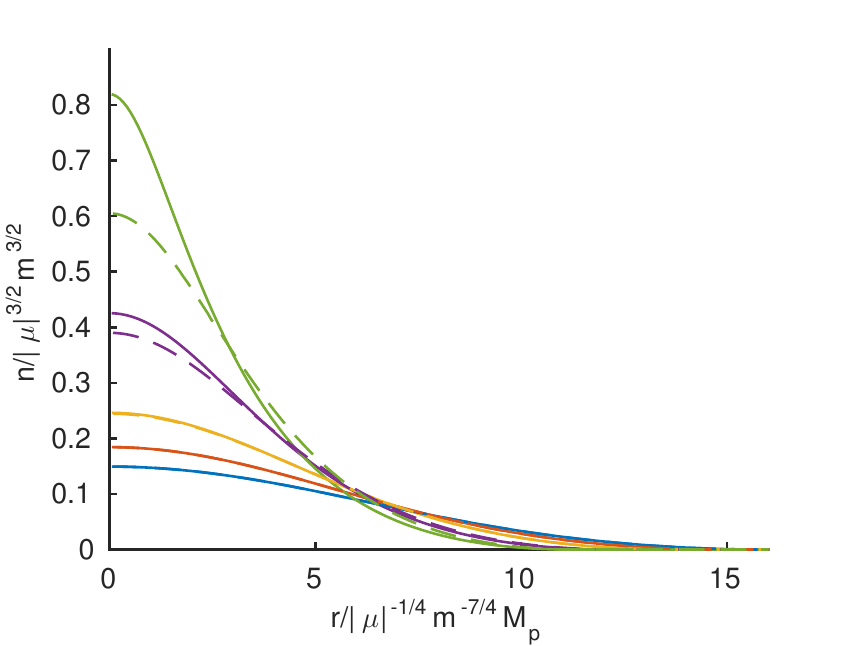}
\caption{\label{fig:gvary}Density profile for different values of coupling constant $g$. The different profiles correspond to $g/(|\mu|^{-1/2}m^{-3/2})=0$ (blue), $1.69$ (red), $3.30$ (yellow), $4.68$ (purple), and $5.11$ (green). The dashed lines show the corresponding results if we fix $\Delta(r)=0$, i.e.\ BCS condensation is assumed not to occur.}
\end{figure}

\begin{figure}
\includegraphics[scale=1]{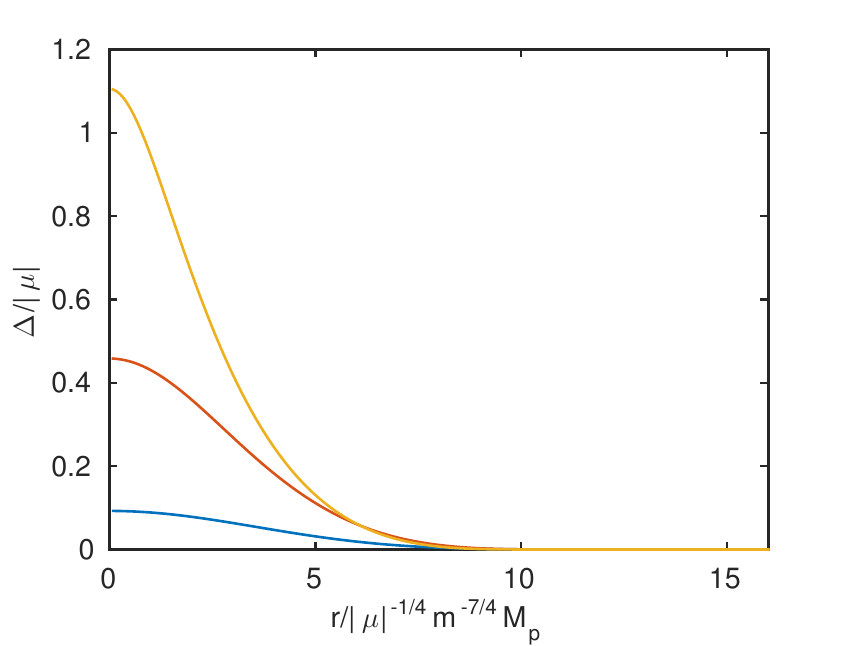}
\caption{\label{fig:gap}Gap energy in units of the chemical potential as a function of radius. Plotted for the three largest values of $g$ from figure \ref{fig:gvary}: $g/(|\mu|^{-1/2}m^{-3/2})=3.30$ (blue), $4.68$ (red), and $5.11$ (yellow).}
\end{figure}

In figures \ref{fig:gvary} and \ref{fig:gap} we plot the density profile and gap energy for different values of the coupling constant, $g$. In figure \ref{fig:gvary} we see that, as we might expect, increasing the attractive interaction strength causes the halo to contract and become more sharply peaked. As the strength of the coupling is increased further, the halo becomes unstable to collapse and no non-trivial self-consistent solution can be found. This puts an upper limit on the possible strength of any attractive coupling between fermions in a dark matter halo.

In figure \ref{fig:gap} we plot the gap energy, $\Delta$ as a function of radius. For the largest coupling strength, $g/(|\mu|^{-1/2}m^{-3/2})=5.11$, the gap energy at the center of the halo becomes comparable to the chemical potential. This coincides approximately with the point at which the halo becomes unstable to collapse due to the attractive interaction. The gap energy goes to zero for somewhat smaller radii than the corresponding density profile. This indicates that outer parts of the halo will not be in the superfluid BCS state.

In figure \ref{fig:gvary} we also plot with a dashed line the density profiles that would occur with no gap energy. This allows us to see how much of an effect the BCS condensation has on the density profile. It is only for the very largest coupling strengths that density profiles with and without the gap become distinguishable. These are the situations where the gap energy becomes comparable to the chemical potential. Most of the contraction of the halo is due to the Hartree interaction term, $gn(r)/2$, with the presence of the gap contributing a further contracting effect for the strongest couplings.

We can now estimate the physical strength of the coupling required to create a significant gap energy and the critical temperature of the corresponding BCS state. We use as a benchmark for a reasonable fit to a dwarf galaxy a fermion mass of $200eV$ \cite{Domcke:2015} and a chemical potential of $|\mu|=10^{-7}\mathrm{eV}$ (see figure \ref{fig:densg0}). The plotted solutions have $g/(|\mu|^{-1/2}m^{-3/2})\sim 1$ which would then require a coupling constant of order $g\sim \mathrm{eV}^{-2}$. Returning to the effective field theory interpretation of the attractive interaction, the associated mass scale would be $M^*\sim\mathrm{eV}$. This is well above the energies in the problem which are of order $|\mu|=10^{-7}\mathrm{eV}$, so treating the interaction as a fermion contact interaction is valid. If torsion is the source of the interaction, the same interaction should occur for all fermions which, at the required strength, would cause issues for electron-quark interactions \cite{Freidel:2005}. Since the interactions in the dark matter halo occur at very low energies, it may be interesting to see whether a running coupling could get around this constraint \cite{Reuter:2015}.

Accurately calculating the critical temperature for the BCS state requires a full finite temperature calculation which is beyond the scope of this work. We can estimate the critical temperature as $T_c\approx \Delta(0)/k_B$ since $\Delta(0)$ is the energy required to produce an excitation above the ground state at the center of the halo. The largest coupling strength for which the halo is stable gives the largest possible critical temperature. From figure \ref{fig:gap} we see that in this case $\Delta(0)\approx |\mu|$. For our benchmark case, $|\mu|=10^{-7}\mathrm{eV}$, this gives a critical temperature of $T_c\sim 10^{-3}\mathrm{K}$. The authors of \cite{Domcke:2015} give an estimate for the temperature of a dwarf galaxy of $10^{-2}\mathrm{K}$ which is not too far away from our estimate of the maximum critical temperature.

In order for a dark matter particle to have a mass as low as $200\mathrm{eV}$ without causing problems for structure formation, it must be produced non-thermally. One method for this is described by Domcke and Urbano \cite{Domcke:2015} where the dark matter is produced directly from decay of the inflaton and is never in thermal equilibrium with the universe. The temperature of such a particle, once it thermalizes with itself, depends on the ratio of the inflaton mass to the reheating temperature. For reasonable values of this ratio ($m_\mathrm{inflaton}\lesssim 10^2T_\mathrm{RH}$), the dark matter can cool sufficiently by the present day for the BCS transition to occur. It will be important in future work to determine precisely when the BCS transition occurs. 

By using the dimensionless variables of equation \eqref{eq:dimless} in equation \eqref{eq:rhocorr} it can be seen that all four terms are suppressed by a factor of $|\mu|/m$ relative to the dominant mass density contribution. For a halo with $|\mu|=10^{-7}\mathrm{eV}$ and $m=200\mathrm{eV}$ this results in a suppression of $10^{-9}$ so we are safe in ignoring these corrections.

\section{Discussion}In this work we have investigated a BCS superfluid non-rotating, spherically symmetric model for the dark matter halo for simplicity. In this case the density profile of the halo, and hence the associated rotation curve, is affected by the presence of the gap energy only for very strong coupling. The presence of the attractive coupling does however affect the density profile more directly by causing the halo to contract. 

The density profiles produced by the superfluid fermions all have cores rather than cusps. That is they converge to a constant density at the center of the halo. In this way they are similar to the commonly used cored profile, the pseudo-isothermal (PI) sphere \cite{deBlok:2009},
\begin{equation}
\rho_\mathrm{PI}(r) = \frac{\rho_0}{1+(r/r_C)^2}
\end{equation}
where $\rho_0$ is the density at the center of the halo and $r_C$ describes the radius of the core region. However unlike the PI profile, the superfluid fermion profile goes to zero at a finite radius, while the PI profile falls off as $r^{-2}$ for large radii. Increasing the strength of the attractive interaction causes the superfluid fermion profile to contract with an increased central density, but it always retains a flat core. The halos produced therefore naturally have cores and a well-defined halo mass.

Adding rotation to the halo opens up the possibility of describing superfluid vortices. Such vortices in a rotating BCS superfluid occur in neutron stars where they are important for explaining glitches in the stars' rotational frequencies \cite{Hoffberg:1970,Pines:1985,DeBlasio:1999,Rezania:2000,Elgaroy:2001,Yu:2003, Buckley:2004,Avogadro:2007}.  Vortices in the halo will form lines through the halo.

We can get an estimate of the size of any vortices present from the BCS coherence length \cite{Tinkham:1996},
\begin{equation}
\xi = \frac{\hbar^2k_F}{\pi m \Delta},
\end{equation}
since the superfluid density should not change on length scales shorter than this. The actual size of the vortices depends on temperature and must be determined numerically by solving the BCS system with rotation, however the coherence length gives a reasonable guide as to the order of magnitude \cite{Gygi:1991}. The condition $|\Delta|\ll\hbar^2k_F^2/m$ is met for all but the largest value of the coupling, so we can use the approximation for the gap, equation \eqref{eq:gapsmall}. This gives an expression for the coherence length
\begin{equation}
\xi = \frac{1}{\pi k_F}\exp\left(\frac{2\pi^2\hbar^2}{mgk_F} \right).
\end{equation}
For $|\mu|\approx 10^{-7}\mathrm{eV}$ and $m=200\mathrm{eV}$, the length scale (in meters) associated with the Fermi momentum at the center of the halo is about
\begin{equation}
\frac{1}{\pi k_F(0)}\approx 5\times 10^{-6}\mathrm{m}.
\end{equation}
For a coupling constant of size $g=0.86|\mu|^{-1/2}m^{-3/2}\approx 0.96\mathrm{eV}$, the exponential factor is approximately $7\times 10^{3}$, giving a coherence length of
\begin{equation}
\xi \sim 10^{-2}\mathrm{m}.
\end{equation} 
The exponential dependence on the coupling constant means that the coherence length can be much larger than this if the coupling constant is small, however in this case the critical temperature is likely to be too low for the BCS state to form.

In order to consider rotating halos, the assumption of spherical symmetry must be relaxed. Axisymmetric models of dark matter halos for dwarf galaxies have previously been considered \cite{Hayashi:2012}. It would be interesting to consider both the global effect of rotation on a BCS condensed halo and the creation of vortex lines in such a situation. 

Berezhiani and Khoury have proposed that bosons in a superfluid state could have non-trivial effective Lagrangians that allow bosonic dark matter to reproduce the rotation curves of MOND \cite{Berezhiani:2015,Berezhiani:2016}. The effective degrees of freedom for the bosonic superfluid are the quasiparticle excitations which have non-trivial dispersion relations. It is possible that our fermionic superfluid may also have quasi-particles which can display similar behaviour. We will consider the quasiparticle spectrum, effects of rotation, and finite temperature effects in future work.

\acknowledgments{We wish to thank Robert Caldwell, Leon Cooper and Justin Khoury for their useful comments.}
\bibliographystyle{JHEP}
\bibliography{BCS_DM_JCAPv4}
\end{document}